# The impact of institutional quality on the relation between FDI and house prices in ASEAN emerging countries

Anh H. Le

*Email: anhlh@chonnam.ac.kr. Department of Economics, Chonnam National University, Gwangju, South Korea;*

This study investigates the relationship between house prices and capital inflows in ASEAN emerging economies, and the impact of institutional quality on that relation. Using a unique balanced panel data set of six emerging countries in ASEAN from 2009 to 2019, we employ various econometric techniques to examine the impact of foreign direct investment (FDI) on the house price index. Our findings indicate a long-run relationship and Granger causality from FDI to the house price index in these markets, and we also find evidence of co-movement between the stock price index and the house price index. Additionally, our results suggest that better institutions reduce the impact of FDI on host country housing markets in the context of ASEAN emerging economies. This is one of the first studies to shed light on the role of institutional quality in the effect of FDIs on housing prices in this region.

**Introduction**

The ASEAN Community, which shares a free trade zone, plays an important role in the Asian economy. As of 2018, the population of ASEAN was 655 million, with 329 million of those being part of the labor force. According to World Development Indicators (WDI) and the author's calculations, this economic region contributed $2,975 billion in GDP, comprising 9% of total GDP in Asia and about 3% of global GDP. One notable aspect of ASEAN is its potential for growth in both production and consumption, thanks to its youthful population (60% of which is under 35 years old)

and abundant natural resources. Since the Global Financial Crisis (GFC), foreign direct investment (FDI) in ASEAN countries has rapidly increased while FDI has declined globally. This is highlighted in Figure 1 and 2, which show a clear trend of increasing FDI net inflows in the ASEAN market (around 3.3 times from 2009 to 2018), while the amount of investment has decreased by around 18% at the global level over the same period.

Of the 11 ASEAN countries, the six emerging markets (ASEAN-6, comprising Thailand, Indonesia, Malaysia, the Philippines, Singapore, and Vietnam) make up around 88% of the population and more than 95% of GDP. These major countries also attract approximately 96% of foreign direct capital inflows to ASEAN countries.

As shown in Figure 3, the real residential index in ASEAN-6 has witnessed an appreciation from the first quarter of 2009 (2009Q1) to the first quarter of 2019 (2019Q1), with an exception of house prices in Vietnam. Previous studies (Algieri, 2013; Égert & Mihaljek, 2007; Ferrero, 2015; Glindro, Subhanij, Szeto, & Zhu, 2011; Zietz, Zietz, & Sirmans, 2008), have identified several variables that can impact house prices, including GDP, CPI, housing credit, interest rates, real exchange rates, and current account deficits.

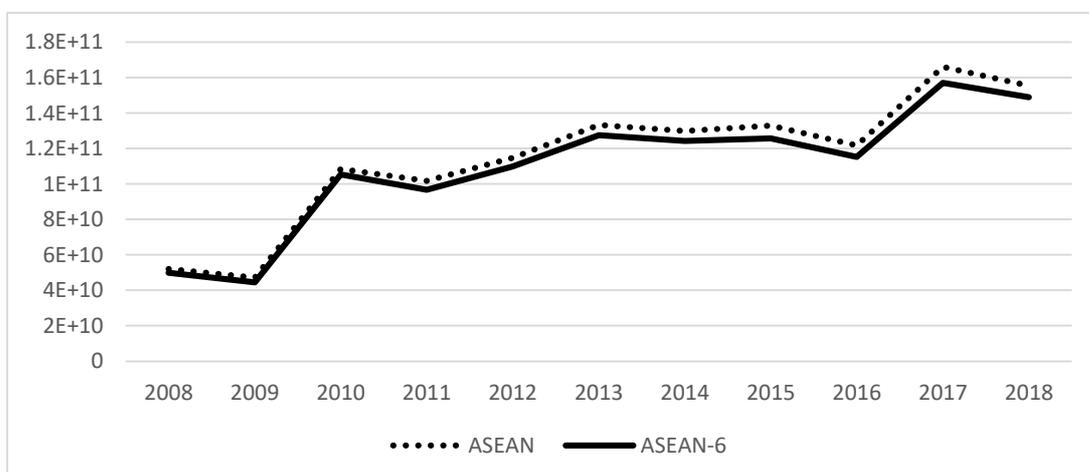

Figure 1 FDI in ASEAN, net inflows (US$). Note. ASEAN-6 includes Singapore, Thailand, Malaysia, Indonesia, Philippines, and Vietnam. Source: World-Bank (2020) and Author's calculation.

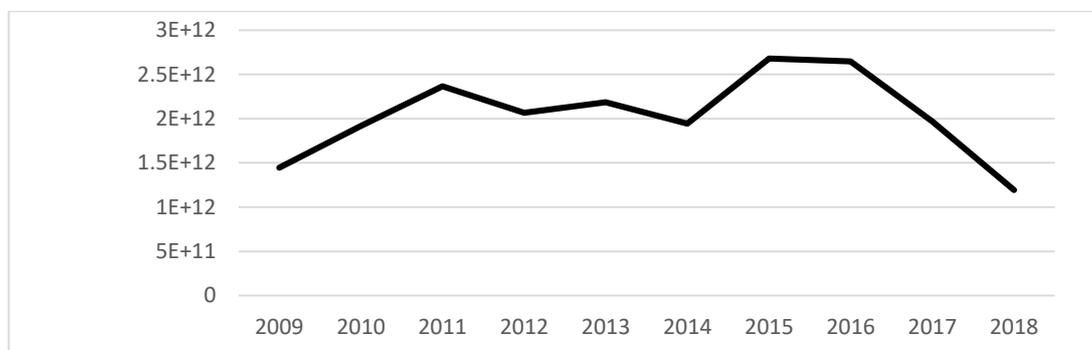

Figure 2 FDI in all the World, net inflows (US$). Source: World-Bank (2020) and Author's calculation.

There has been much interest among scholars in the relationship between capital flows and house prices (Favilukis, Kohn, Ludvigson, & Van Nieuwerburgh, 2012; Tillmann, 2013; Yiu & Sahminan, 2015). Many studies have used current account deficit as a measure of capital inflow, while others have focused on portfolio investment to examine speculation in real estate. In addition, using 35 cross-city panel data in China, Zheng, Kahn, and Liu (2010) found that house prices are lower in cities with high levels of pollution, and that FDI inflows can improve air quality. Ferrero (2015) conducted a study on house prices in advanced and emerging economies from 1990 to 2012 and found that house prices are more closely related to capital flows in emerging countries than in advanced countries. Foreign direct investment and foreign portfolio investment have different characteristics that may affect housing prices differently. Previous research has tended to focus on portfolio investment rather than direct investment. This study aims to fill this gap by examining the impact of foreign direct investment on the house price index in ASEAN-6 countries. We predict that there will be a relationship between house prices and foreign direct investment in these markets.

Table 1 provides a summary of specific studies on the relationship between capital flows and property assets. However, to date, only a limited number of studies have investigated the impact of FDI on house prices in all six emerging ASEAN countries.

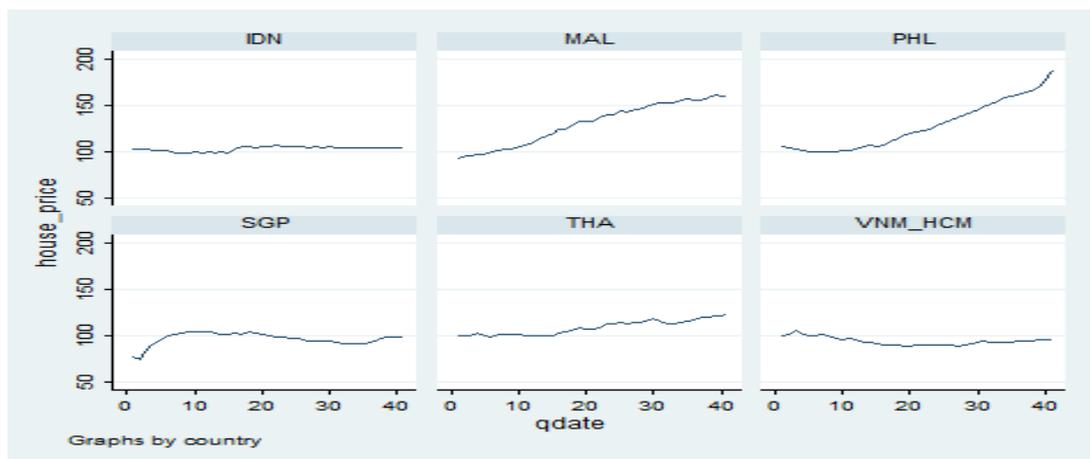

Figure 3 House price in ASEAN-6 from 2009Q1 to 2019Q1. Note: House price index is a real residential property index collected from BIS for Indonesia, Malaysia, Philippines, Singapore, Thailand, and Savills property price index from Savills research and consultancy for Vietnam data.

Foreign direct investment and foreign portfolio investment have different characteristics that may affect housing prices differently. Previous research has tended to focus on portfolio investment rather than direct investment. This study aims to fill this gap by examining the impact of foreign direct investment on the house price index in ASEAN-6 countries. We predict that there will be a relationship between house prices and foreign direct investment in these markets.

Table 1 Brief of literature review of five studies on the relation between capital flows and property assets

| Authors | Period | Country | Methodology | Main variables | Control variables | Highlighted Results |
|---|---|---|---|---|---|---|
| Tillmann (2013) | Quarterly data | South Korea, Malaysia, Taiwan, Thailand, HongKong | Panel VAR approach | Total net capital inflows, Foreign Portfolio investment | Real GDP, Price level | Capital inflows shocks have a significant effect on the appreciation of house prices. |

| | VAR 1: 2000Q1 to 2011Q1 | | | House price | REER appreciation Long/short-run interest rate | |
|---|---|---|---|---|---|---|
| | VAR 2: 2003Q1 to 2010Q4 | HongKong, South Korea, Taiwan, Singapore, Malaysia, Thailand | | | | |
| Kole and Martin (2009) | Quarterly data and Annual data 1970-2009 | United States, United Kingdom, Canada, Germany, France, Japan, Australia, Netherlands, Spain, Sweden | Jointly OLS Country fixed effects US recession | Change in house price Nominal current account to nominal GDP | GDP growth Change in the long term interest rate | The growth rate of house price and current account balance has a negative correlation. Consumption of housing and consumption of market goods are complementary. |
| Aizenman and Jinjarak (2009) | Annual data 1990-2005 | 43 countries (25 OECD countries) | OLS GMM Fixed effect (least squares dummy variable) | Percentage change per year of real estate prices divided by GDP deflator. Current account deficits divided by GDP. | Capita GDP growth, Real interest, Urban population growth, Inflation, Institution, Financial depth, | Current account deficits and the appreciation of real estate prices divided by GDP deflator has a strong positive linkage. |
| Jinjarak and Sheffrin (2011) | Quarterly data Up to 2009 Q1 | Ireland, England, Spain and the United States | VARs Granger causality Graph theory (TETRAD'S PC ALGORITHM) | Current account divided by GDP. Real estate prices. | Interest rates, equity prices, and output growth | Increasing capital inflow directly cause the appreciation of real estate price. (Evidence found in Ireland, Spain, and the United States) |
| Kim and Yang (2011) | Quarterly data 1999Q1-2006Q1 | 5 Asian economies: South Korea, Indonesia, Thailand, Malaysia, and the Philippines. | Panel VAR | Land price Stock price Capital inflows/Portfolio inflows (a ratio to GDP) | Real GDP GDP deflator Short-term interest rate, REER, capital outflows. | Capital inflow shocks have little impact on asset prices. |

Note: (1) (YYYY)**Q**(#) denote quarter (#) of the year (YYYY). (2) REER denotes real effective exchange rate. (3) VAR denotes Vector autoregression. (4) OLS is Ordinary Least Square and GMM is General Method of Moments. (5) OECD is the Organization for Economic Co-operation and Development.

Foreign investment inflows can also impact the domestic real estate industry. An increase in the number of expats may lead to an increase in demand for residential property. In particular, in developing countries, FDI projects can bring improvements in infrastructure and facilities, which can support an increase in expected residential prices. On the other hand, FDI may create pollution in the host country, which can lower the quality of life and therefore decrease expected residential prices. The finding from this step lays the groundwork for further investigating the impact of institutional quality on the relationship between foreign direct investment and house prices.

The direct impact of institutional quality on the real estate market has been investigated in several articles (Glindro, Subhanij, & Szeto, 2011; Levin & Satarov, 2000; Musole,

2009; Pan, Huang, & Chiang, 2015; Toulmin, 2009). However, there is a lack of studies that have looked at the impact of institutions on the relationship between foreign capital flows and domestic house prices. We hypothesize that improvements in institutional quality can provide incentives and relax financing constraints for domestic investors, leading to more dynamic activity among domestic investors and weakening the impact of foreign capital investment on house prices.

To test this hypothesis, we have constructed a unique balanced panel including data from six emerging countries in ASEAN from 2009Q1 to 2019Q1. The house price index is a real residential property index collected from the Bank for International Settlements (BIS) for Indonesia, Malaysia, the Philippines, and Singapore, and from Savills property price index and research consultancy for Vietnam. Quarterly data on FDI is collected from the CEIC database. Other explanatory variables are obtained from reliable sources such as the International Monetary Fund (IMF), World Bank (WB), Heritage Foundation, and the General Statistics Office of Vietnam (GSO).

This research will have important policy implications. The recent Global Financial Crisis (GFC) was closely linked to the real estate market and the occurrence of housing bubbles. Despite this, ASEAN has successfully attracted a large amount of foreign investment flows to the region through its economic freedom and improvements in property rights. However, the potential risk of housing bubbles accompanying high volumes of foreign capital flows remains. Studying the long-run and short-run correlation between FDI and house prices can provide policymakers with valuable evidence.

The rest of this paper is structured as follows: Section 2 describes the variables and data used in the study; Section 3 outlines the methodology; Section 4 presents the results and discussion; and Section 5 provides the conclusion.

**Data descriptive**

Our key variables are *HP* and *FLOW*. First, *HP* is a natural logarithm of the house price index. Second, *FLOW* is a natural logarithm of foreign direct investment inflow. Data sources are mentioned in the first section of this paper. The house price index is a real residential property index collected from BIS (2020a) for Indonesia, Malaysia, Philippines, Singapore, Thailand, and Savills property price index from Savills-research-and-consultancy-for-Vietnam (2020). Quarterly data of FDI is collected from CEIC (2020).

Our control variables are *INCOME, INTEREST, EXRATE, STOCKPRICE*, and *INST*, defined as follows. *INCOME* measures by a natural logarithm of nominal GDP adjusted by inflation, follow the calculation as $INCOME = Ln(\frac{Nominal\ GDP}{CPI} \times 100)$. Where nominal GDP is quarterly nominal gross domestic product and CPI is quarterly consumer price index. Vietnam's normal GDP collected from GSO-Vietnam (2020) and convert to USD by the official exchange rate collected from IMF/IFS (2020). Data for nominal GDP in other countries is quarterly collected from CEIC (2020). *CPI* is quarterly obtained from IMF/IFS (2020).

*INTEREST* is a lending interest rate. Data is quarterly available from IMF (2020). *EXRATE* is a natural logarithm of the exchange rate against USD. Quarterly data is available from BIS (2020b).

*STOCKPRICE* is a natural logarithm of the quarterly average stock index. Quarterly data is converted from daily price. The daily price is obtained from Wall-Street-Journal (2020).

*INST* is the index of economic freedom. This is annual data obtained from Heritage-Foundation (2020). This variable is used to measure the role of institutional quality on the relation between FDI and house price.

Table 1 shows the descriptive statistics of the main variables for the whole sample and each country. On average, house price indices in Malaysia and the Philippines are higher than the whole sample figure and also more volatile than in other countries. Singapore attracts the largest amount of capital flow, holds the strongest currencies but the house price index is lower than the whole sample. Vietnam is the poorest country still absorbed many FDI inflows. Indonesia and Vietnam hold the weakest currencies and have a very high lending rate.

Table 1 Descriptive statistics of main variables

|  | *HP* | *FLOW* | *INCOME* | *INTEREST* | *EXRATE* | *STOCKPRICE* | No. Obs |
|---|---|---|---|---|---|---|---|
| All panels | 4.69 | 7.50 | 11.1863 | 0.07 | 4.69 | 7.31 | 246 |
|  | 0.17 | 2.99 | 0.55 | 0.03 | 3.72 | 1.18 |  |
| Indonesia | 4.63 | 7.93 | 12.1023 | 0.12 | 9.33 | 8.35 | 41 |
|  | 0.02 | 2.79 | 0.123 | 0.01 | 0.18 | 0.34 |  |
| Malaysia | 4.86 | 7.01 | 11.1340 | 0.05 | 1.27 | 5.45 | 41 |
|  | 0.18 | 3.22 | 0.123 | 0.00 | 0.13 | 0.15 |  |
| Philippines | 4.83 | 6.57 | 10.9723 | 0.06 | 3.83 | 8.63 | 41 |
|  | 0.20 | 2.12 | 0.148 | 0.01 | 0.07 | 0.39 |  |
| Singapore | 4.57 | 9.61 | 11.1199 | 0.05 | 0.29 | 8.01 | 41 |
|  | 0.07 | 0.46 | 0.134 | 0.00 | 0.05 | 0.13 |  |
| Thailand | 4.69 | 6.04 | 11.4377 | 0.05 | 3.48 | 7.11 | 41 |
|  | 0.07 | 4.87 | 0.119 | 0.00 | 0.05 | 0.34 |  |
| Vietnam | 4.54 | 7.82 | 10.3518 | 0.10 | 9.94 | 6.34 | 41 |
|  | 0.05 | 0.30 | 0.208 | 0.03 | 0.08 | 0.30 |  |

Note: (1) This table reports the summary statistics of key variables, in each country and the whole sample, from quarter 1 2009 to quarter 1 2019 (2009Q1-2019Q1). (2) For each variable, the numbers in the first row represent the sample mean and those in the second row represent the standard deviation. (3) HP denotes the house price index, FLOW denotes foreign direct inflow, INCOME is nominal gross domestic product adjusted by inflation, INTEREST is interest rate, EXRATE is exchange rate against US dollar, STOCKPRICE is stock index. All variables (except INTEREST) are in natural logarithm form.

To consider the impact of institutional quality, this essay uses the index of economic freedom released by the Heritage Foundation. They classify as 5 zones. The first class is free with a score of over 80. Second, the mostly free zone has a score between 70 and 79.9. Third, the moderately free zone has a score between 60 and 69.9. Forth, the mostly unfree zone has a score between 50 and 59.9. Fifth, the repressed zone has a score of less than 50.

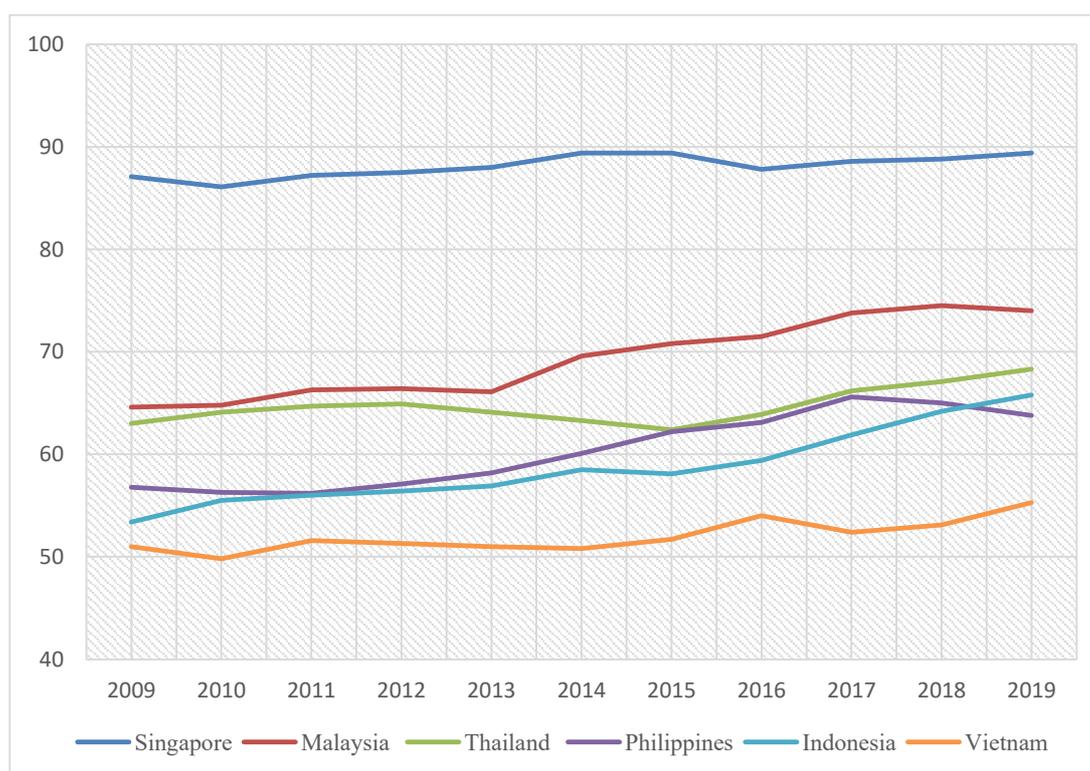

Figure 1 Index of economic freedom for ASEAN-6 obtained from Heritage-Foundation (2020).

Figure 1 illustrates the overall score of the index of economic freedom for ASEAN-6

from 2009 to 2019. Over the research period, Singapore is ranked in the first class of freedom with a very high score, whereas Vietnam appears at the lowest score among ASEAN-6 and is still classified as the mostly unfree environment. Malaysia, the Philippines, and Indonesia have dramatically witnessed an improvement in their economic environment, and jump to a higher zone. Thailand was in the moderately free zone.

**Methodology**

Assume the long-run house price function,

$$HP_{it} = \alpha_i + \beta X_{it} + \mu_{it} \quad (4)$$

$$X_{it} = [\ FLOW_{it}\ INCOME_{it}\ INTEREST_{it}\ EXRATE_{it}\ STOCKPRICE_{it}\ ]$$

In the majority of unit root tests, house price index, and explanatory variables stationary at *I(1)*. Thus, this study will follow a framework as bellows.

Firstly, this essay employs panel cointegration tests introduced by Kao (1999), and Pedroni (1999, 2004). Secondly, Dumitrescu and Hurlin (2012) Granger non-causality test is employed to test panel causality. Thirdly, panel fully modified ordinary least square (FMOLS) is used to study the relationship between house price and capital flows. FMOLS allows different causality relations between house prices and capital inflows across countries. However, FMOLS only estimates the long-run relations. Thus this essay will employ pool mean group model (PMG) estimation to shed some light on the short-run relation. Also, PMG approach allows variables of a different order of stationarity. After these tests, we use panel fixed effect regression to the panel error-correction models (ECM). Similar to many previous authors, such as (Le & Kim, 2020), we augment the model with an interactive term of institution quality and capital flows to investigate the role of institutional quality on the relationship between capital flows and house price index.

**Panel fixed-effect regression**

We use panel regression with panel and time fixed effect to check the role of institutional quality. The equation is given as follows.

$$D.HP_{it} = \sigma_i * (HP_{i,t-1} + \beta_i * X'_{i,t-1})) + \sigma_i * D.HP_{it} + \beta_i * D.X'_{i,t-1} + \varepsilon_{it} \quad (9)$$

$$where\ X' = [FLOW, INST\_FLOW, INTEREST]^1$$

**Results** and **discussions**

*Panel cointegration analysis*

Firstly, panel cointegration tests are conducted. This step is applied Kao's tests and Pedroni's tests for cointegration. Kao (1999) and Pedroni (1999, 2004) developed the idea of Engle and Granger (1987). Assume two data series $X_t$ and $Y_t$ are stationary at *I(1)*. If $X_t$ and $Y_t$ have cointegrated relation, the residual of $X_t$ and $Y_t$ is stationary at *I(0)*. Kao (1999) provides five tests to conduct the null hypothesis (H0), no cointegration, and an alternative hypothesis (Ha), all panels are cointegrated. Then Pedroni (1999, 2004) contribute to the previous author's work with several tests. More importantly, they allow heterogeneous panels.

Table 1 Kao's cointegration test

| Modified Dickey-Fuller t | -1.9884** |
|---|---|
| Dickey-Fuller t | -1.5803* |
| Augmented Dickey-Fuller t | -0.5423 |

---

[1] We try other control variables, but they show insignificantly statistics.

| Unadjusted modified Dickey | -1.2826* |
| --- | --- |
| Unadjusted Dickey-Fuller t | -1.2992* |

Note: (1) *$p < 0.1$, **$p < 0.05$, ***$p < 0.01$. (2) This table shows a Kao test for cointegration with the null hypothesis (H0), no cointegration, and an alternative hypothesis (Ha), all panels are cointegrated. (3) Cointegrating vector is the same, included panel means in the model for dependent variable on the covariates, autoregressive parameter as the same for all panels. (4) The number of panels is 6, and the number of periods is 39. (5) The method used to estimate the long-run variance of each panel's series (Kernel type) is Bartlett. (6) Lags are 3.0 (Newy-West) and augmented lags is 1. (7) Dependent variable is house price index (*HP*), and independent variables are foreign direct investment inward (*FLOW*), nominal gross domestic adjusted by inflation (*INCOME*), interest rate (*INTEREST*), exchange rate against US dollar (*EXRATE*), quarterly average stock index (*STOCKPRICE*), all variables (except *INTEREST*) in the natural logarithm form.

Table 1 and Table 2 report the cointegration test statistics. In Table 1, except Unadjusted Modified Dickey-Fuller, other tests reject the null hypothesis. The first test Dickey-Fuller t rejects the null hypothesis at the 5% significance level. The other three tests, Modified Dickey-Fuller t Augmented Dickey-fuller t and Unadjusted Dickey-Fuller t, reject the null hypothesis at the 10% significance level. In Table 2, Modified Variance ratio, Modified Phillips-Perron t reject the null hypothesis at 10%, 5%, and 1% significance levels, respectively. The majority of Kao's tests and Pedroni's test supports the existence of the cointegration relation between the house price index and our explanatory variables.

Table 2 Pedroni's cointegration test

| Modified variance ratio | -3.6929*** |
| --- | --- |
| Modified Phillips-Perron t | 2.9072*** |
| Phillips-Perron t | 0.6853 |
| Augmented Dickey-Fuller t | -1.1516 |

Note: (1) *$p < 0.1$, **$p < 0.05$, ***$p < 0.01$. (2) This table shows a Pedroni test for cointegration with the null hypothesis (H0), no cointegration, and an alternative hypothesis (Ha), all panels are cointegrated. (3) Cointegrating vector is panel specific,

included panel means and panel-specific linear time trends in the model for dependent variable on the covariates, autoregressive parameter as the same for all panels. (4) The number of panels is 6, and the number of periods is 40. (5) The method used to estimate the long-run variance of each panel's series (Kernel type) is Bartlett. (6) Lags are 3.0 (Newy-West) and augmented lags is 1. (7) Dependent variable is house price index (*HP*), and independent variables are foreign direct investment inward (*FLOW*), nominal gross domestic adjusted by inflation (*INCOME*), interest rate (*INTEREST*), exchange rate against US dollar (*EXRATE*), quarterly average stock index (*STOCKPRICE*), all variables (except *INTEREST*) in the natural logarithm form.

*Panel Causality analysis*

In the following step, this essay applies Dumitrescu and Hurlin (2012) Granger non-causality test to investigate the direction of causality between FDI and house price index. The null hypothesis (H0) is that *FLOW* does not Granger-cause *HP*, and the alternative (Ha) is *FLOW* does Granger-cause *HP* for at least one panel. Table 3 reports that there is a long-run Granger causality running from *FLOW* to *HP* at least one panel. This means that FDI inflows do Granger-cause house price change at least in one country.

Table 3 Granger non-causality tests

| | |
|---|---|
| W-bar = | 4.3117 |
| Z-bar = | 2.8312*** |
| Z-bar tilde = | 2.3677 ** |
| Lag order | 2 |

Note: * $p < 0.1$, ** $p < 0.05$, *** $p < 0.01$. This table presents the test for Granger non-causality from foreign direct investment inward (FLOW) to house price index (HP) in heterogeneous panels using the procedure proposed by Dumitrescu and Hurlin (2012).

*Panel fully modified ordinary least square*

Table 4 reports the panel FMOLS tests with pooled estimation and group estimation. All coefficients are positive and significant at least 5% level with an exception of the coefficient of STOCKPRICE under pooled estimation.

Table 4 Panel FMOLS

| Variable | Pooled estimation | | Grouped estimation | |
|---|---|---|---|---|
| | Coefficient | t-Statistic | Coefficient | t-Statistic |
| FLOW | 0.0080 | 2.11** | 0.0176 | 2.89*** |
| INCOME | 0.3478 | 3.33*** | 0.2050 | 3.85*** |
| INTEREST | 2.6388 | 3.17*** | 14.1920 | 3.04*** |
| EXRATE | 0.4831 | 4.20*** | 0.3207 | 6.81*** |
| STOCKPRICE | 0.0947 | 1.59 | 0.1475 | 3.42*** |

Note: * $p < 0.1$, ** $p < 0.05$, *** $p < 0.01$. This table presents the FMOLS results for pooled estimation and grouped estimation. The pooled estimator simply sums across cross-sections, while a grouped-mean estimator takes averages over the individual cross-sections. Dependent variable is house price index (HP), and independent variables are foreign direct investment inward (FLOW), nominal gross domestic adjusted by inflation (INCOME), interest rate (INTEREST), exchange rate against US dollar (EXRATE), quarterly average stock index (STOCKPRICE), all variables (except INTEREST) in the natural logarithm form.

First, the long-run coefficients of *FLOW* are positive at 5% and 1% significance level in Pooled estimation and grouped estimation, respectively. In the long run, higher FDI inflows likely accelerate the house price. Second, the coefficients of *INCOME, INTEREST*, and *EXRATE* are positive at 1% significance level in both estimations. The positive coefficients of *INCOME* provide evidence for the income's effect on house prices. Reasonably, an improvement in income follows with an increase in demand for houses.

The positive correlation between house prices and interest rates can be explained by the hedging perspective. On one side, the lending rate negatively affects the demand side. As house purchases are usually financed by mortgage credits, a high-interest rate raises the cost to purchase a house. Consequently, house buyers are discouraged to join the housing market. House price decreases because of lower demand. On the opposite side, the interest rate implies expected inflation in the coming period (Eugene F. Fama, 1975; Eugene F Fama & Schwert, 1977). From the hedges' point of view, housing is the main asset against the depreciation of the currency. With a positive correlation between INTEREST and HP, our result gives strong evidence for hedging purposes based on investor's expectations of inflation. Similarly, the exchange rate against USD measures the relative value of domestic currency and USD. Consistent with previous studies, house price and exchange rate have a positive correlation.

Third, the coefficient of *STOCKPRICE* is only significant under grouped FMOLS estimation. The figure is positive at 1% significance level. A co-movement of stock price and house price is supported in our result. According to Glindro et al. (2011), the stock price has two opposite effects on house price, the substitution effect, and the wealth effect. Our results show that the wealth effect is greater than the substitution effect. Besides, the stock price index may also provide a reliable signal for the business cycle. Investors buy and sell stocks basing on their expectations of the economic condition in the coming period. Therefore, if they expand their investment in the stock market, they likely invest in other assets such as real estate.

***Pool mean group estimator***

Table 5 shows the long-run coefficients. With the exception of *EXRATE* and *INTEREST*, other explanatory variables have positive correlations with *HP* at least 5% significance level. Note that the coefficient of *INTEREST* turns negative in the PMG estimator, but it is not still significant statistically. Other variables are consistent with the result of the FMOSL approach.

Table 5 PMG estimation results (*Long run equation)*

| Variables | Coefficient (t-statistics) |
|---|---|
| FLOW | 0.0122** |
|  | (2.06) |
| INCOME | 0.4281*** |
|  | (4.48) |
| INTEREST | -2.6179 |
|  | (-1.56) |
| EXRATE | 0.0807 |
|  | (0.55) |
| STOCKPRICE | 0.1746*** |
|  | (3.50) |

Note: *t* statistics in parentheses, $^*p < 0.1$, $^{**}p < 0.05$, $^{***}p < 0.01$. This table presents the PMG estimation result for long-run correlations between house price index (*HP*), and independent variables, which are foreign direct investment inward (*FLOW*), nominal gross domestic adjusted by inflation (*INCOME*), interest rate (*INTEREST*), exchange rate against US dollar (*EXRATE*), quarterly average stock index (*STOCKPRICE*), all variables (except *INTEREST*) in the natural logarithm form.

Table 6 shows the negative figures and significant statistics of the speed of adjustment to long-run equilibrium ($\phi$). This result supports the jointly long-run cointegrating relation between *HP* and the group of explanatory variables. $\phi$ is showed at the highest figure in Singapore and the lowest figure in Philippines. This table also reports the short-run relations between *HP* and *FLOW*. Some coefficients are significant but all of them are very small, which implies a very weak relation between *HP* and *FLOW* in the short run.

Table 6 PMG estimation results (*Short-run equation*)

| Panels/Variables | COINTEQ01 | D(HP(-1)) | D(FLOW) | D(FLOW(-1)) | D(FLOW(-2)) | D(FLOW(-3)) |
|---|---|---|---|---|---|---|
| Total | -0.1641** | -0.0204 | -0.0018 | -0.0006** | 0.0005 | -0.0019 |
|  | -2.11 | -0.11 | -1.43 | -2.39 | 0.27 | -1.66 |
| Indonesia | -0.0472*** | -0.0137 | -0.0009*** | -0.0004*** | -0.0002*** | 4.14E-06*** |
|  | -129.59 | -0.70 | -5550.57 | -2426.16 | -1654.05 | 46.34 |
| Malaysia | -0.0342*** | -0.3803*** | -0.0017*** | -0.0014*** | -0.0017*** | -0.0011*** |
|  | -20.88 | -10.69 | -5912.39 | -3774.73 | -4229.74 | -4243.47 |
| Philippines | -0.0291*** | 0.2622*** | -0.0009*** | -0.0005*** | 7.09E-05*** | -0.0001*** |
|  | -131.05 | 7.56 | -3163.39 | -1161.78 | 141.71 | -496.06 |
| Singapore | -0.5122*** | 0.7743*** | 0.0021*** | -0.0014*** | 0.0094*** | -0.0041*** |
|  | -54.83 | 42.46 | 92.88 | -49.03 | 405.09 | -256.94 |
| Thailand | -0.2529*** | -0.3460*** | -0.0021*** | -0.0006*** | 3.37E-05*** | 0.0004*** |
|  | -63.58 | -21.45 | -1821.69 | -811.93 | 114.23 | 2608.75 |
| Vietnam | -0.1090*** | -0.4190*** | -0.0076*** | 0.0004 | -0.0045*** | -0.0065*** |
|  | -33.52 | -26.36 | -41.88 | 1.71 | -28.62 | -104.12 |

Note: (1) For each country the number in the first row is the coefficient and those in the second row represent the t statistics. (2) *$p < 0.1$, ** $p < 0.05$, *** $p < 0.01$. (3) This table presents the PMG estimation result for the speed of adjustment to long-run equilibrium ($\phi$), and short-run correlations between house price index (*HP*), and independent variables, which are foreign direct investment inward (*FLOW*). (4) COINTEQ01 denotes the speed of adjustment to long-run equilibrium ($\phi$). D(X(-#)) denotes difference operators of variable *X* at level #.

*The role of institutional quality*

Table 7 Panel fixed effect regression – the role of institutional quality

|  | (1) D.HP | (2) D.HP |
|---|---|---|
| L.HP | -0.0207** | -0.0259** |
|  | (-2.06) | (-1.99) |
| L.FLOW | 0.00130** | 0.00132** |
|  | (2.14) | (2.17) |
| L.INST_FLOW | -0.00191*** | -0.00204*** |
|  | (-3.03) | (-3.08) |
| L.INTEREST | -0.195*** | -0.163* |
|  | (-2.63) | (-1.82) |
| LD.HP | 0.186*** | 0.192*** |
|  | (2.88) | (2.94) |
| LD.FLOW | -0.000528 | -0.000541 |
|  | (-1.28) | (-1.31) |
| LD.INST_FLOW | 0.000742 | 0.000809* |
|  | (1.64) | (1.74) |
| LD.INTEREST | 0.755** | 0.767*** |
|  | (2.58) | (2.61) |
| Time fixed-effect | No | Yes |
| N | 234 | 234 |

Note: z statistics in parentheses, $^{*}$ $p < 0.1$, $^{**}$ $p < 0.05$, $^{***}$ $p < 0.01$. This table presents the panel regression result for equation (9). Dependent variable is natural logarithm of house price index (*HP*), and independent variables are natural logarithm of foreign direct investment inward (*FLOW*), interest rate (*INTEREST*), INST_FLOW is the interaction term between *FLOW* and economic freedom index (INST). L. means the first lag of variable and D. means difference operator of variable.

In Table 7, the coefficients of *L.FLOW* are positive and significant at the *p = 0.05* level, whereas the coefficients of *L.INST_FLOW* are negative and significant at the *p = 0.01* level. This result implies that a better economic environment reduces the house price acceleration followed by the high volume of FDI flows. Improvement in the index of economic freedom indicates an improvement in property rights, financial markets, and control of corruption, which strengthens the domestic investment environment and reduces the sensitivity of domestic price to foreign capital inflows.

**Conclusions**

Appreciation of house prices is a key concern in emerging markets. The close relationship between the mortgage credit market and the current global financial crisis has prompted many scholars to study housing price bubbles. This study contributes to the existing literature on this topic by examining the impact of foreign direct investment on house prices and the role of institutional quality in ASEAN-6 countries. As liberalization has progressed, many countries in ASEAN have successfully attracted high volumes of foreign direct capital inflows and experienced rapid growth in their housing markets. Using various techniques such as FMOLS, PMG, and panel regression with fixed effects, this study finds evidence of a positive long-run relationship between house prices and FDI inflows. However, the effect of FDI flows on house prices is negatively impacted by improvements in the index of economic freedom.

This research has important implications for policymakers in ASEAN-6 countries. We suggest that these countries should consider adjusting their laws and regulations to support more economic freedom. With better institutions, the domestic market may be less sensitive to foreign intervention.

**Acknowledgment**

I would like to express my gratitude to Professor Jaewoon Koo and Professor Taegi Kim for their valuable advice and support throughout this research project. Their guidance and expertise were invaluable in helping me to develop and refine my ideas, and I am deeply grateful for their contributions.